\documentclass[11pt,preprint]{aastex}

\def\ltsima{$\; \buildrel < \over \sim \;$}
\def\lsim{\lower.5ex\hbox{\ltsima}}
\def\gtsima{$\; \buildrel > \over \sim \;$}
\def\gsim{\lower.5ex\hbox{\gtsima}}
\def\mpl{$M_{\rm P}$}
\def\rpl{$R_{\rm P}$}
\def\shat{$\hat{\mathbf s}$}
\def\nhat{$\hat{\mathbf n}$}
\def\khat{$\hat{\mathbf k}$}

\slugcomment{Accepted for publication in the Astrophysical Journal {\it Letters}}
\shorttitle{Obliquity tides}
\shortauthors{Winn and Holman}

\begin{document}

\title{ Obliquity Tides on Hot Jupiters }

\author{
Joshua N.\ Winn\altaffilmark{1} and Matthew J.\ Holman
}

\affil{Harvard-Smithsonian Center for Astrophysics, 60 Garden Street,
Cambridge, MA 02138}

\altaffiltext{1}{Hubble Fellow}

\begin{abstract}
Obliquity tides are a potentially important source of heat for
extrasolar planets on close-in orbits. Although tidal dissipation will
usually reduce the obliquity to zero, a nonzero obliquity can persist
if the planet is in a Cassini state, a resonance between spin
precession and orbital precession. Obliquity tides might be the cause
of the anomalously large size of the transiting planet HD~209458b.
\end{abstract}

\keywords{stars:\ individual (HD~209458)---planets and
satellites---planetary systems: formation}

\section{Introduction}

A wonderful surprise of the last decade was the discovery of ``hot
Jupiters'': giant planets that orbit their parent stars at a distance
smaller than 10 stellar radii (Mayor \& Queloz 1995, Butler et al.\
1997). The proximity of the star and planet raises the possibility of
tidal interactions between them. Of these interactions, the most
frequently discussed are the synchronization of the planetary rotation
period and orbital period, which happens over $\sim$$10^6$~yr, and
orbital circularization, over $10^{8-9}$~yr (see, e.g., Rasio et al.\
1996, Lin \& Gu 2004). Less frequently discussed, but often implicit,
is tidal evolution of the obliquity (the angle between the planetary
spin axis and the orbit normal), which occurs on the same time scale
as synchronization (Peale 1999).

We wish to point out that in some circumstances a close-in planet can
maintain a nonzero obliquity, and that the consequent heat from
obliquity tides can be large enough to affect the internal structure
of the planet. We present the heating calculation first, in \S~1,
since it does not depend upon our particular scheme for maintaining
the obliquity. Our proposal, given in \S~2, is that hot Jupiters
occupy Cassini states, in which spin precession resonates with orbital
precession. In \S~3 we ask whether obliquity tides can explain the
famously small density of the extrasolar planet HD~209458b. In \S~4,
we discuss the implications, strengths, weaknesses, and possible tests
of this theory.

\section{The power of obliquity tides}

Tidal torques synchronize the planetary spin frequency ($\omega$) and
orbital mean motion ($n$), and often reduce the obliquity ($\theta$)
to zero (Goldreich \& Peale 1970). Henceforth, if the orbit is
circular, the tidal bulge is motionless in the reference frame of the
planet and energy dissipation ceases. However, dissipation continues
if the planet somehow maintains a nonzero eccentricity or
obliquity. For a star of mass $M_\star$ and a planet of mass \mpl\ and
radius \rpl\ in an orbit with semimajor axis $a$, the rate of energy
loss through tidal friction is
\begin{equation}
\frac{dE}{dt} = \left[
\frac{9}{10} \frac{h}{Q}
n
\left( \frac{GM_{\rm P}^2}{R_{\rm P}} \right)
\left( \frac{M_\star}{M_{\rm P}} \right)^2
\left( \frac{R_{\rm P}}{a} \right)^6
\right] 
\left( 7 e^2 + \sin^2 \theta \right).
\label{eq:heat}
\end{equation}
This formula, derived in a different context by Wisdom (2004), employs
the customary model of friction within astronomical bodies: $Q$ is the
``quality factor'' of tidal oscillations (the inverse of the
fractional energy dissipated per cycle). The factor $h$ is the
displacement Love number, parameterizing our ignorance of the planet's
deformability. The factor in square brackets may be written
\begin{equation}
\left[
2\times 10^{27} \hspace{0.03in} {\rm erg~s}^{-1} \hspace{0.05in}
\left( \frac{Q/h}{10^{6}} \right)^{-1}
\left( \frac{P}{{\rm 3~days}} \right)^{-5}
\left( \frac{R_{\rm P}}{R_{\rm Jup}} \right)^5
\right],
\end{equation}
where $P$ is the orbital period.

For this heat source to play a significant role in determining the
structure of the planet, it must be comparable to the intrinsic
luminosity $L_0$ of other processes (e.g.\ gravitational contraction,
stellar irradiation, radioactivity, and nuclear reactions). For
Jupiter, $L_0 = 3\times 10^{24}$~erg~s$^{-1}$. For hot Jupiters,
theoretical estimates of $L_0$ are difficult to summarize (depending
as they do upon the planet's age, temperature, composition,
atmosphere, and interior structure) but for billion-year-old planets
of Jupiter's radius and mass, one expects $L_0\approx 2\times
10^{25}$~erg~s$^{-1}$ (see, e.g., Guillot \& Showman 2002, Baraffe et
al.\ 2003). Tidal heating is dominant when $e\gsim 0.04$ or
$\theta\gsim 0.1$. The importance of eccentricity tides is well known
(Bodenheimer, Lin, \& Mardling 2001), but the equivalent importance of
obliquity tides does not seem to have been appreciated.

\section{Cassini states for extrasolar planets}

The reason is probably a mistaken intuition that $\theta=0$ is the
only possible endpoint of tidal evolution. In general, the planet's
spin axis and its orbit both precess, in response to additional
planets, satellites, the stellar quadrupole, or other torquing
agents. With spin and orbital precession, the outcomes of tidal
evolution are Cassini states (Colombo 1966, Peale 1969, Ward 1975), in
which the orbit normal \nhat\ and spin axis \shat\ precess at the same
rate about the same axis \khat\ (see Fig.~1). For a given body, there
are at most two stable Cassini states, differing in whether \shat\ and
\nhat\ are on the same side of \khat\ (state 1) or opposite sides
(state 2).\footnote{Formally, there are two other equilibria: state 3
is linearly stable but is unstable to tidal evolution (Goldreich \&
Peale 1970), and state 4 is linearly unstable.} The obliquity $\theta
= \cos^{-1} (\hat{\mathbf s} \cdot \hat{\mathbf n})$ of a Cassini
state is not generally zero, nor is it necessarily close to zero.
Although many satellites in the Solar system are in state 1 with
vanishingly small obliquity, the Moon is in state 2 with
$\theta=6\fdg5$.\footnote{The observation that the Moon's \shat,
\nhat, and \khat\ are coplanar was the basis of G.~D.\ Cassini's third
law of lunar motion, formulated in 1693.}

\begin{figure}[h]
\epsscale{1.0} \plotone{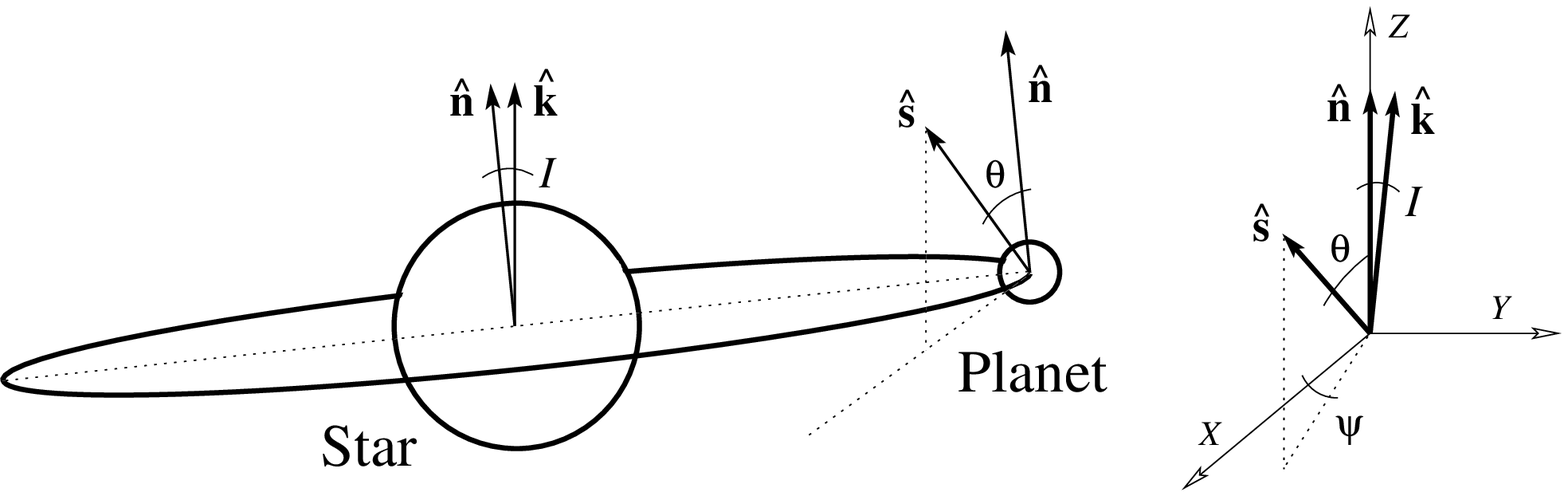}
\caption{({\it Left}.) Illustration of the spin axis \shat, orbit
normal \nhat, precession axis \khat, orbital inclination $I$ and
obliquity $\theta$. A general orientation for \shat\ is depicted. In a
Cassini state, \shat\ is coplanar with \khat\ and \nhat. ({\it
Right}.) A convenient set of Cartesian axes, in which \nhat\ is along
the $Z$ axis, \khat\ is in the $YZ$ plane, and $(\theta,\psi)$ are
traditional polar angles describing \shat. In a Cassini state,
$\psi=\pm 90\arcdeg$.
\label{fig:diagram}}
\end{figure}

Idealizing the planet as an oblate rigid body on a circular and
uniformly precessing orbit of inclination $I = \cos^{-1}(\hat{\mathbf
k}\cdot\hat{\mathbf n})$, the Cassini obliquities $\theta_i$ obey
\begin{equation}
\cos\theta_i \sin\theta_i - \epsilon \sin(\theta_i - I) = 0.
\label{eq:cassini-obliquity}
\end{equation}
Here we have defined $\epsilon = -g / \alpha$, where $g$ is the nodal
precession frequency and $\alpha$ is the spin precessional constant
for a fixed orbit. The latter can be written
\begin{equation}
\alpha = \frac{3}{2} \left( \frac{C-A}{C} \right)
\left( \frac{n^2}{\omega} \right),
\label{eq:alpha}
\end{equation}
where $C>B=A$ are the planet's principal moments of inertia, and
$\omega$ is its spin frequency.  During synchronization,
$\omega\rightarrow n$.  When $\epsilon < \epsilon_{\rm crit} \equiv
(\sin^{2/3} I + \cos^{2/3} I)^{-3/2}$,
Eq.~(\ref{eq:cassini-obliquity}) has 4 roots, corresponding to the two
stable states, 1 and 2, and two unstable states, 3 and 4. For
$\epsilon > \epsilon_{\rm crit}$, there are only 2 roots,
corresponding to states 2 and~3. Eq.~(\ref{eq:cassini-obliquity}) can
be derived from the governing Hamiltonian under the assumption of
principal-axis rotation (see, e.g., Ward 1975),
\begin{equation}
{\mathcal H} = {\mathcal H}_0 -\frac{\alpha}{2}
( \hat{\mathbf n}\cdot \hat{\mathbf s} )^2 -
g ( \hat{\mathbf k}\cdot \hat{\mathbf s} ),
\label{eq:hamiltonian}
\end{equation}
where ${\mathcal H}_0$ is the \shat-independent portion. Fig.~2 shows
contours of ${\mathcal H}$(\shat) for an illustrative case.

\begin{figure}[h]
\epsscale{1.0}
\plotone{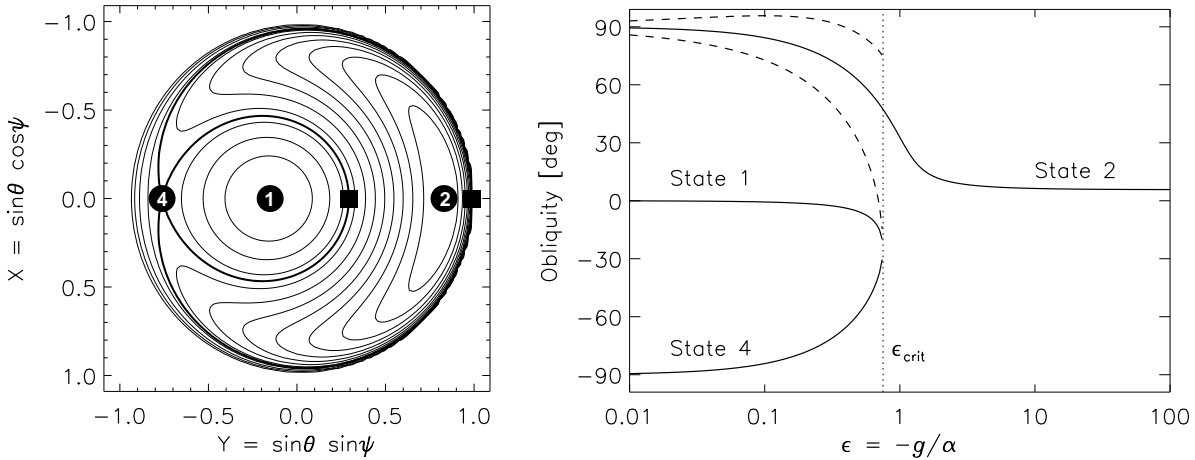}
\caption{({\it Left.}) Contours of ${\mathcal H}$(\shat), for
$\epsilon=0.6$ and $I=0.1$. The unit sphere is projected onto the $XY$
plane. The thick contour is the separatrix, which divides the basins
of attraction of states 1 and 2 under tidal evolution. The numbered
dots show states 1, 2, and 4. The retrograde state 3 is not
shown. The separatrix angles $\theta_{\rm s}$ are marked by
squares. ({\it Right.}) Cassini state obliquities $\theta_i$ (solid
lines) and separatrix angles $\theta_{\rm s}$ (dashed lines) as a
function of $\epsilon$, for $I=0.1$.
\label{fig:hamiltonian}}
\end{figure}

A hot Jupiter settles into a Cassini state on the same $\sim$$10^6$~yr
time scale as spin-orbit synchronization. Whether it ends up in state
1 or 2 depends upon its initial obliquity and orbital inclination, and
upon $\epsilon$, which may be a function of time. Reasons for
variations in $\epsilon$ include the disappearance of the
protoplanetary disk, migration of the planet or its fellow planets or
satellites, contraction of the planet, and spin alteration of the
planet or star. If $\epsilon$ varies slowly compared to $g^{-1}$ and
$\alpha^{-1}$ (``adiabatically''), and slowly compared to the Cassini
state settling time, then the obliquity tracks the evolving Cassini
obliquity. Abrupt changes in $\epsilon$ or $\theta$ may cause the
planet to leave a Cassini state, and ultimately to switch states if it
lands in the basin of attraction of the other stable state.

Fig.~2 suggests three ways in which a hot Jupiter can maintain a
significant obliquity. First, if $\epsilon > \epsilon_{\rm crit}$, the
only stable state is 2, for which the obliquity is nonzero
($\theta_2\rightarrow I$ as $\epsilon \rightarrow \infty$). This is
the case for the Moon. Second, it could be in state 1 or 2 with
$\epsilon\sim \epsilon_{\rm crit}$, but this requires a special
coincidence, since $\epsilon$ involves both planet-specific properties
(namely, its moments of inertia) and seemingly unrelated quantities
specific to other bodies (which determine the precessional
torques). Third, the planet could be in state 2 with $\epsilon <
\epsilon_{\rm crit}$. When $\epsilon$ is small, the basin of
attraction of state 2 is also small, but it is possible to be captured
into state 2 when $\epsilon$ is large and then tipped over to large
obliquity as $\epsilon$ is gradually reduced. Ward \& Hamilton (2004)
proposed this scenario to explain why Saturn has a much larger
obliquity than Jupiter. In the next section, we propose a similar
scenario for a particular extrasolar planet.

\section{The case of HD~209458b}

The planet HD~209458b transits its parent star (Charbonneau et al.\
2000, Henry et al.\ 2000), a fortuitous circumstance that enables many
interesting measurements, including that of the planet's mean density,
which is 0.33~g~cm$^{-3}$. This is 27\% of the Jovian value and is the
smallest mean density of all 7 known transiting extrasolar planets
(see, e.g., Alonso et al.\ 2004, Konacki et al.\ 2005, Pont et al.\
2005). Theorists have struggled to explain this anomaly, usually by
attempting to identify an overlooked internal heat source, although
Burrows, Sudarsky, \& Hubbard (2003) argued that the density is not so
terribly anomalous. We present a new hypothesis: HD~209458b resides in
Cassini state 2 with a large obliquity, whereas most hot Jupiters
reside in state 1 with small obliquities.

We are led to imagine the following sequence of events: (1) The planet
forms at a large orbital distance, with a nonzero (but not necessarily
large) obliquity. (2) The planet migrates inward to its current
position. (3) As the spin and orbit are synchronized over
$\sim$$10^6$~yr, the planet falls into Cassini state 2, whether by
chance or because state 2 is the only possibility. (4) As the disk
disappears over $\sim$$10^7$~yr, the orbital precession rate (and
$\epsilon$) decreases, forcing the obliquity to grow. (5) The planet
remains in state 2 for billions of years.

Is this scenario compatible with order-of-magnitude estimates of the
relevant quantities? We assume $I=0.1$, because Winn et al.\ (2005)
found an angle of $\approx$4\arcdeg\ between the sky projections of
the stellar spin axis and orbit normal, and it seems reasonable that
$I$ is of the same order of magnitude. We estimate $(C-A)/C$ by
scaling Jupiter's observed value (de Pater \& Lissauer 2001) by
$(\omega/\omega_{\rm Jup})^2$. Application of Eq.~(\ref{eq:alpha})
gives $\alpha/n \sim 10^{-3}$ after synchronization. We assume that
$g$ was initially determined by the torque of the protoplanetary disk,
which we idealize as a minimum-mass solar nebula. The precession rate
is sensitive to the (unknown) inner radius of the disk: for $r_{\rm
min} = 1.6 a$, $|g/n| = 5\times 10^{-5}$, rising to $10^{-4}$ for $1.3
a$. These estimates ignore resonance effects, which we find through
trial numerical integrations to be capable of enhancing $g$ by an
order of magnitude. As the disk disappears, $g$ falls until other
torques dominate, such as those from the stellar quadrupole or
additional orbiting bodies. In the former case, assuming $J_2 =
10^{-6}$, the precession slows to $|g/n| = 4\times 10^{-8}$. There is
no evidence for additional bodies in current radial velocity data
(Laughlin et al.\ 2005), but even a body too small or too distant to
have been detected could nevertheless cause faster precession than the
stellar quadrupole alone. For example, a 1$M_\oplus$ planet in a
1.2~day orbit would cause $|g/n| = 2\times 10^{-5}$, while producing
undetectable radial velocity variations of $\sim$0.5~m~s$^{-1}$. We
note that $g$ is measurable, in principle, through time variations of
the transit duration (Miralda-Escud\'{e} 2002).

With these considerations, it seems plausible that $\epsilon$ had an
initial value ranging from 0.02 to near unity, and a final value
between $10^{-2}$ and $10^{-5}$. The initial value of $\epsilon$ could
have been large enough that capture into Cassini state 2 occurred with
reasonable probability, or even 100\% probability if
$\epsilon>\epsilon_{\rm crit}$. The final value of $\epsilon$ is small
enough to force $\theta_2\rightarrow 90\arcdeg$. Thus, the current
obliquity must be quite large.

The energy dissipated through obliquity tides comes at the expense of
the orbital energy. We must examine whether it is possible for the
dissipation rate to be large enough to inflate the planet, yet slow
enough to allow the orbit to survive for $5\times 10^9$~yr, the
approximate main-sequence age of the star (Cody \& Sasselov 2002). To
inflate the planet, Bodenheimer, Laughlin, \& Lin (2003) found the
required power to be $4\times 10^{27}$~erg~s$^{-1}$ if the planet has
a dense core, and an order of magnitude smaller if the planet is
coreless. With reference to Eq.~(\ref{eq:heat}), this corresponds to
an upper limit on $Q/h$ of $10^6$ for a cored planet and $10^7$ for a
coreless planet. The condition
\begin{equation}
\frac{a}{|da/dt|} = \frac{G M_\star M_{\rm P} / 2a} {dE/dt}
> 5\times 10^9 \hspace{0.05in} {\mathrm{yr}},
\label{eq:timescale}
\end{equation}
yields a lower limit on $Q/h$ of $5\times 10^6$. In reality,
Eq.~(\ref{eq:timescale}) is probably too restrictive by a factor of a
few; the dissipation rate was smaller in the past, when the planet had
a smaller radius, a larger orbital distance, and possibly a larger
mass (Vidal-Madjar et al.\ 2003). Thus, $Q/h$ should be $\sim$10$^6$
if the planet has a core and $10^6$--$10^7$ if it does not. In
comparison, for Jupiter it is thought that $Q$ is between $10^5$ and
$10^6$ (Goldreich \& Soter 1966) and $h\gsim 0.6$ (Gavrilov \& Zharkov
1977). The required $Q/h$ for HD~209458b is comparable to, or somewhat
larger than, the nominal Jovian value.

The last chapter of the story is that the planet remains in Cassini
state 2 for billions of years. Here the difficulty is that as
$\epsilon$ decreases, the width of the resonance also decreases,
reducing the robustness of state 2 to perturbations. For
$\epsilon=10^{-5}$, the angle between the separatrix angles (see
Fig.~2) is only $0\fdg2$. For $\epsilon=10^{-2}$, it is increased to
$7\arcdeg$. The impact of a body of mass $m$ at escape velocity would
produce a maximum obliquity shift of
\begin{equation}
\frac{mv_{\rm esc} R_{\rm P}}{L_\omega} =
\frac{m\sqrt{2GM_{\rm P}R_{\rm P}}}{Cn} =
20\arcdeg \left(\frac{m}{M_\oplus}\right),
\end{equation}
where the maximum is achieved for grazing incidence at the pole.
Hence we must also suppose HD~209458b suffered no major collisions
after the disappearance of the protoplanetary disk.

\section{Discussion}

Hot Jupiters should be in Cassini states. Since the obliquity of a
Cassini state is not necessarily small, obliquity tides are a
potentially important internal heat source. Whether a given planet can
maintain a significant obliquity depends upon its initial obliquity as
well as its precessional and collisional histories. Obliquity tides
could be the ``missing'' heat source that bloats the transiting planet
HD~209458b. The implications of this hypothesis are that the planet's
obliquity is nearly 90\arcdeg, its tidal dissipation factor $Q$ and
displacement Love number $h$ obey $10^6 < Q/h < 10^7$, and it had a
quiescent history of no major collisions after its orbital precession
rate declined from an initially large value.

This hypothesis has certain strengths and weaknesses relative to the
two leading prior hypotheses. Guillot \& Showman (2002) proposed that
atmospheric circulation patterns convert a small fraction of the
stellar radiation into heat deep within the planet. An advantage of
this hypothesis over ours is that no fine tuning of $Q/h$ or the
collision history is required. A disadvantage is that it is not
obvious why the other hot Jupiters should not experience the same
phenomenon. In contrast, Cassini state 2 is naturally a minority state
for hot Jupiters. After landing in state 2 (possibly by chance), a
planet must avoid being jostled into state 1 by collisions or the
synchronization process.

In the scenario of Bodenheimer et al.\ (2001), tidal heating from
orbital circularization is extended indefinitely because of a periodic
eccentricity exchange with another planet. The extra heat is the fault
of a well-placed third body, which is naturally expected to be a rare
occurrence. Unfortunately, no such body has yet been detected, and the
current eccentricity is small (Laughlin et al.\ 2005, Winn et al.\
2005). These problems have made the hypothesis less appealing than it
once was, though they might be overcome through tuning of $Q/h$ or
other embellishments. A relative strength of our hypothesis that no
third body is required, although a third body could increase the
robustness of the high-obliquity state.

Further work is needed to improve upon our order-of-magnitude
calculations. The synchronization process must be followed in detail
because $\alpha$ changes on the same time scale as obliquity
evolution, and is therefore non-adiabatic (Peale 1974). We treated the
planet as an oblate spheroid, but in reality it is triaxial due to
tidal distortion, a complexity that alters the Cassini obliquities and
dynamics (Peale 1969). We ignored any evolution of $e$ or $I$, and any
non-uniform precession. Finally, a realistic description of
perturbations is needed to estimate the capture probability and
lifetime of Cassini state 2.

We can think of two possible tests of the theory that HD~209458b has a
large obliquity, neither of which is easy. First, the obliquity can be
measured or bounded through high-precision photometry of the transit
(Seager \& Hui 2002, Barnes \& Fortney 2003), but the expected signal
is smaller than $10^{-5}$ in relative flux. Second, a hot Jupiter with
$\theta\approx 90\arcdeg$ has no permanent dayside or nightside, and
should have a smaller day--night temperature difference than a planet
with $\theta=0\arcdeg$. To quantify the expected temperature
difference, atmospheric models of hot Jupiters (Showman \& Guillot
2002, Cho et al.\ 2003, Menou et al.\ 2003, Burkert et al.\ 2005)
should be generalized to cases of large obliquity.

\acknowledgments We are indebted to J.\ Wisdom for his suggestion to
investigate Cassini states. We thank R.\ Crocker, S.\ Gaudi, G.\
Laughlin, D.\ Lin, S.\ Peale, K.\ Penev, D.\ Sasselov, K.\ Stanek, and
W.\ Ward for helpful discussions. Work by J.N.W.\ was supported by
NASA through grant HST-HF-01180.02-A.

\end{document}